\documentstyle[12pt]{article}
\newcommand{\be}{\begin{equation}}
\newcommand{\ee}{\end{equation}}
\newcommand{\ba}{\begin{eqnarray}}
\newcommand{\ea}{\end{eqnarray}}

\newcommand{\la}[1]{\label{#1}}

\def\gl#1{(\ref{#1})}

\begin{document}
\centerline{{\large\bf Spontaneous CPT asymmetry of the 
Universe}\footnote{Talk at V International Conference on Cosmoparticle
Physics COSMION-2001, dedicated to the 80-th Anniversary of Andrei D. 
Sakharov (Moscow-St. Petersburg, May 21-30, 2001)}} 
\vspace{0.5cm}

\centerline{A.A.Andrianov$^{1,2}$, P.Giacconi$^1$, R.Soldati$^1$}
\centerline{$^1$ Dip. di Fisica \& INFN, Univ. di Bologna, via Irnerio 46, 
40126, Bologna}
\centerline{$^2$ Dept. of Theor. Phys., V. A. Fock Inst. of Physics, 
198504, 
St.Petersburg}
\vspace{0.5cm} 

{\footnotesize The extended QED with renormalizable interactions breaking 
the
Lorentz and CPT symmetry is considered and the phenomenological consequences of 
such a symmetry breaking are illuminated in view of recent
discussion of large scale anisotropy of the Universe.
Other physical effects in QED with CPT violation are examined, 
in particular, mass splitting between electrons of different
helicities and decay of very high energetic electrons  into lower energy
electrons and positrons.}
\vspace{0.5cm}

So far the Lorentz symmetry has been proven to hold
 with a very high accuracy. Nevertheless, one can inquire about
 whether the Special Relativity Theory is 
 only approximate. The modern, quantum field-theoretical approach
 admits that the {\it spontaneous} Lorentz symmetry breaking is not
 excluded. At this expense the CPT symmetry can be also broken in a
 local field theory.

 The occurrence of a very small deviation from the Lorentz
invariance has been discussed recently [1-3], within the
context of the Standard model of electroweak interactions. There,
some "background" or "cosmological" fields are implied , leading
to deviations in the propagation of certain
particles, within the present experimental limits. 

As the photon is a test particle of the Special Relativity the
most crucial probes concern  Lorentz- and
CPT-symmetry breaking modifications of electrodynamics.

When one retains its fundamental character provided by
renormalizability and the gauge invariance of its action, 
then it might be induced in the ordinary {\it 3+1}
dimensional Minkowski space-time by supplementing QED with the
additional CPT-odd Chern-Simons (CS) coupling of photons to the
vacuum [1] mediated by a constant CS vector $\eta_\mu$.
 This modification of Electrodynamics
 does not break the gauge symmetry of the action but changes the
dispersion relations for different photon polarizations [1].
As well a constant axial-vector field $b_\mu$ may couple to the fermions [2]
splitting their masses and breaking the CPT and Lorentz symmetry.

All together the minimal modification of QED is given by the
following lagrangian,
\be
{\cal L} =-\frac14 F_{\mu\nu}F^{\mu\nu} - B n_\mu A^\mu - \frac12 \eta_\mu
A_\nu \tilde F^{\mu\nu}
 + \bar\psi\left(i\gamma^\mu \partial_\mu  + e \gamma^\mu A_\mu - m
- \gamma^\mu \gamma _5 b_\mu\right) \psi \label{CPTodd}
\ee
in the axial gauge. The Lorentz and CPT symmetry breaking is
parameterized by two vectors $\eta_\mu$ and $b_\mu$ which are not
necessarily collinear. Their dynamical origin represents an
interesting problem and
one of the possible ways to induce LSB by a dynamical mechanism
has been suggested recently [4]. Namely,
 the spontaneous breaking of the
Lorentz symmetry {\it via} the Coleman-Weinberg mechanism [5] has
been proved for a class of models with the Wess-Zumino interaction
between abelian gauge fields and pseudoscalar massless axion $\theta(x)$(AWZ
models). Then the condensation of the gradient of axion field
$<\partial_\mu \theta> \sim  \eta_\mu \sim b_\mu $ may
be a natural implementation for LSB [6] just relating its origin
to the existing ``quintessence'' fields [7]. On the other hand, a
part of the background vector $b_\mu$ may represent the constant torsion
$b_\mu \sim \epsilon_{\mu\nu\rho\sigma} T^{\nu\rho\sigma}$ in the 
large scale Universe [8-10].

When having the modified QED \gl{CPTodd} one searches  for the
phenomenological consequences of the symmetry breaking which can
be registered in the experiments. As well one should examine
the consistency of this QFT (stability [4,11,12] and microcausality[11,13])
once the latter one pretends to describe the
particle physics in a wide range of energies.\\

\underline{\large\it Distorted photons}\\

In the CPT odd QED the {\it linearly} polarized photons created in the
Universe may reveal the birefringence [1], {\it i.e.} the rotation of
the polarization direction depending on the distance, when they
propagate in the vacuum. The consistent QED exists only for 
{\it space-like} CS vectors $\eta^2 < 0$ (in the presence of fermions
 one has also to take into account
the radiatively induced CS term [12,14]) and it can be canonically quantized
in the frames where this vector is entirely space-like, $\eta_0 = 0$
(see[4,12,13]).
In this case there are always four real solutions of the equation for
energy spectrum
\be
k^2_{0\pm}({\vec k},\vec\eta) =
{\vec k}^2+{1\over 2}\vec\eta^2\pm
|\vec\eta|\sqrt{(\vec k\cdot\vec\eta)^2+{1\over 4}\vec\eta^2}\ .
\ee
As it can be seen for very small wave vector $\vec k$ two photon modes 
(the ``left'' ones) remain massless and other two (the ``right'' ones) become massive. Therefore
monochromatic plane wave
solutions are possible only with a definite chiral polarization
whereas
the linearly polarized photons reveal the rotation of the polarization
direction in time = with a distance.
We stress that  in the case of CPT odd QED 
the notion of handedness or chirality which is conserved
 only approximately corresponds to that one in the Maxwell QED, {i.e.}
the left-(right-)handed photons do not have exactly circular polarizations
[1,2,15].
Meanwhile the group velocities for these modes
turn to be always smaller than $c$.
This supports a
birefringence phenomenon, because the group velocities of the linearly 
polarized wave-packets
made out of chirally polarized CS-photons are smaller than $c$.

The earlier analysis of radiowaves from distant galaxies had been
performed [1] in the
assumption of purely time-like LSB  when $\vec\eta = 0$ which however
is not consistent at the quantum level. The space-like scenario was recently
reexamined [16,17]. The present situation can be conservatively characterized
by an upper bound, $|\vec\eta|\leq 10^{-32} eV$. \\

\underline{\large\it Fermions in the constant axial field}\\

The consistent Dirac-type quantization of massive fermions in the
constant axial-vector background is frame-dependent and available for
{\it time-like} $b^2 > 0$ if the space part is small $\vec b^2 < m^2$.
In this case one has four real solutions: two of them with positive energies and 
other two with negative energies [12].
In the rest frame of $b = (b_0,0,0,0)$ their dispersion law is
\be
p_0 = \pm\sqrt{(|\vec p| \pm |b_0|)^2 + m^2};\ \ p^2 =  b_0^2 + m^2 \pm 2 |b_0||\vec p| .
\la{dis}
\ee
One can see that a pair of solutions of a definite helicity always describes the massive particle
but another pair of opposite helicity corresponds to a massive particle only
for $|\vec p| <  (b_0^2 + m^2) /2 |b_0|$. Above this bound the phase velocity exceeds
the conventional speed of light $c$. Nevertheless if  $b^2 > 0$ then the group velocity
turns out to be $< c$. 

Thus in such a theory the two characteristic scales arise:
one is supposedly small $|b_\mu| \ll m$ and another one is extremely high
$M \simeq m^2/2|b_0| \gg m$. Respectively two phenomena occur at low and high
energies. The first one is the mass splitting between  electrons of different
helicities ($\pm$ in \gl{dis}). 
if the value of $b_0$ is universal for all fermions, then 
the lightest massive ones - electrons 
and positrons -  should give the best precision in its determination.
As the precision of a measurement of the electron mass
[18] is of the order $10^{-8}$, the upper bound on the
time component of CPT breaking
vector is not very stringent: namely, $|b_0| < 10^{-2}\ {\rm eV}$ 
(or even weaker, being controlled by the energy
resolution in accelerator beams).
On the other hand,  some much more stringent
bounds were obtained for space components of $b_\mu$, from 
the experiments with atomic systems using
hydrogen masers [19] - {\it i.e.} $|\vec b| < 10^{-18}\ {\rm eV}$ - 
and with a spin-polarized torsion pendulum [20] - {\it i.e.}
$|\vec b| < 10^{-20}\ {\rm eV}$.

The second phenomenon takes place at very 
high energies, $M > m^2/|b_0|\simeq 10^2 TeV$ when some of electron states
obtain the space-like energy-momentum. Then conventional wisdom of relativity
theory fails and a high energetic electron 
 may well decay into an electron of the same helicity
and a pair of positron and electron of the positive and negative helicities
respectively.
This process starts for $|\vec p| > 2 m^2/|b_0| $ and 
the final momenta of particles
are roughly three times less than the initial momentum. Such a process will
``wash out'' from the asymptotic states very high energetic electrons and positrons and
therefore the natural momentum cutoff arises being stuck to the
rest frame of the vector $b_\mu$.  The ultraviolet divergences are then
cured by this physical cutoff and the effective CS vector radiatively 
induced by one fermion loop is defined uniquely [12],
independently on renormalization scheme (compare to [14,21]),
\be
\Delta\eta_\mu = {2\alpha \over\pi}\sum_{a=1}^N b^{a}_\mu,
\la{rad}
\ee
where the summation has to be performed over all the $N$ internal
charged fermions degrees of freedom and the possibility to have different
axial charges for different fermions is taken into account.From
the recent experimental bounds obtained in Refs.~[19,20]
it is possible to estimate the magnitude of the spatial components 
of the induced CS vector to be 
of the order $|\Delta{\vec \eta}|<10^{-19}\ {\rm eV}$, under the
assumption that all the
vectors ${\vec b}^{a}$ are of the same order.
If the vectors $b^{a}_\mu$ are related to some background torsion,
then one expects them to be identical.
If, however, they are generated
by {\it e.g.} vacuum expectation values of gradients of axion fields,
then it is conceivable that $b^{a}_\mu$ might have different values.\\

\underline{\large\it Consistency between photons and fermions}\\

The consistent quantization of fermions in a constant
axial-vector field asks for the vectors $ b^{a}_\mu$ to be time-like 
[11, 12] which, nevertheless, does
not mean that the sum in eq.~\gl{rad} is also time-like, because some of the
fermions may have the opposite axial hypercharges.
 
On the other hand, the consistent quantization of photons can be achieved
when the full {\sl dressed} CS vector
$\tilde\eta_\mu = \eta_\mu + \Delta\eta_\mu$
turns out to be  essentially purely space-like [12,13].

As we suppose that there
is some dynamical mechanism to generate -
with the help of axion condensation -
the purely bosonic part $\eta_\mu$ of the full CS four-vector,
the compatibility of the
consistent quantization of both fermions and bosons is believed to be
quite possible, contrary to the claim in Refs.~[13].

However, from the practical point of view, the cancellation between 
time components
of  $\eta_\mu$  and $\Delta\eta_\mu$ should be extremely
precise, in order to satisfy the experimental bounds [1,16,17] and
to fulfill the microcausality requirement of the photo-dynamics.
As well the severe experimental bounds on  $\vec b^{a}$ for electrons,
and protons [19,20], together with the estimation of birefringence
of radio-waves from remote galaxies and quasars[1,16,17], do not leave too much
room to eventually discover the CPT and Lorentz symmetry breaking
in the quantum spinor and photon dynamics.

\vspace{0.5cm}

A. A. A. is partially supported by Grant RFBR
01-02-17152, Russian
Ministry of Education Grant E00-33-208 and by The Program {\it Universities
of Russia: Fundamental Investigations} (Grant 992612); P. G. is supported
by Grant MURST-Cofin99.\\

%

\end{document}